\documentclass[aps,
               prb,
               reprint,
               showpacs,
               showkeys,
               amsmath,
               amssymb,
superscriptaddress]
              {revtex4-1}

\usepackage[utf8]{inputenc}
\usepackage[T1]{fontenc}
\usepackage{times}
\usepackage{float}

\usepackage[colorlinks=true,
            citecolor=blue,
            urlcolor=blue,
            linkcolor=blue]
           {hyperref}

\usepackage{csquotes}

\usepackage{mathtools}
\DeclarePairedDelimiter{\abs}{\lvert}{\rvert}
\DeclarePairedDelimiter{\ket}{\lvert}{\rangle}

\DeclarePairedDelimiterX{\ketbra}[2]{\lvert}{\rvert}{#1\delimsize\rangle\delimsize\langle #2}
\DeclarePairedDelimiterX{\expval}[3]{\langle}{\rangle}{#1\delimsize\vert #2\delimsize\vert #3}

\DeclarePairedDelimiter{\set}{\{}{\}}

\newcommand{\vcorrect}[2]{\ensuremath{\lower\dimexpr#1-\fontdimen22\textfont2\relax\hbox{#2}}}

\begin{document}

\title{Double Semion Phase in an Exactly Solvable Quantum Dimer Model \\ on the Kagome Lattice}

\author{Oliver Buerschaper}
\affiliation{Perimeter Institute for Theoretical Physics,
    31 Caroline Street North,
    Waterloo,
    Ontario,
    Canada,
    N2L 2Y5}
\author{Siddhardh C. Morampudi}
\author{Frank Pollmann}
\affiliation{Max-Planck-Institut für Physik komplexer Systeme,
    Nöthnitzer Straße 38,
    01187 Dresden,
    Germany}

\begin{abstract}
    Quantum dimer models typically arise in various low energy theories like those of frustrated antiferromagnets.
    We introduce a quantum dimer model on the kagome lattice which stabilizes an alternative $\mathbb{Z}_2$~topological order, namely the so-called \enquote{double semion} order.
    For a particular set of parameters, the model is exactly solvable, allowing us to access the ground state as well as the excited states.
    We show that the double semion phase is stable over a wide range of parameters using numerical exact diagonalization.
    Furthermore, we propose a simple  microscopic spin Hamiltonian for which the low-energy physics is described by the derived quantum dimer model.
\end{abstract}

\maketitle

\section{Introduction}

Topologically ordered phases of matter currently attract a lot of attention as they represent a fundamentally new form of matter, which cannot be classified by symmetry breaking and also exhibit very unusual properties.\cite{WEN1990}
One of their most interesting properties is the existence of emergent anyonic quasi-particle excitations which obey fractional statistics, i.e. they are neither bosons nor fermions.\cite{Leinaas1977,Wilczek1982}
Proposals to use these anyons as a building block for a robust topological quantum computer have pushed forward interest in topologically ordered phases.\cite{Kitaev2003,Nayak2008}
There exists a wide variety of such phases whose understanding comes from their low energy theories which are topological field theories.
These phases are characterized by the statistics of the anyonic quasi-particles which is often summarized in the $U$- and $S$-matrices.\cite{KESKI-VAKKURI1993}

An important question is where to find physical systems that stabilize such phases.
The most successful approaches so far are based on fractional quantum Hall systems which have been shown to realize various types of topological order at different filling fractions.\cite{Laughlin1983,WenFQH1990,Nayak2008}
Another fertile ground for the realization of topologically ordered states are frustrated magnets.
In such systems, the geometry of the lattice prohibits a simultaneous minimization of all the interactions between the spins, leaving some of them frustrated.
This frustration can then destabilize conventional orderings, opening the possibility to form disordered liquid-like ground states.
For example, it is currently debated whether the spin-$1/2$ Heisenberg model on the kagome lattice forms a topologically ordered \enquote{spin liquid}.\cite{Yan2011,Jiang2012,Depenbrock2012}

In the context of frustrated systems, so-called quantum dimer models (\textsc{QDM}) play an important role as effective low-energy descriptions.\cite{Rokhsar:1988ko,Misguich:2003bd,Poilblanc2010}
These \textsc{QDM}s were originally proposed in the context of resonating valence-bond (\textsc{RVB}) states in the theory of high-temperature superconductors.\cite{AndersonRVB1987}
Rokhsar and Kivelson later constructed a simple \textsc{QDM} Hamiltonian which can be fine tuned to a particular point (\textsc{RK}~point) for which the ground state is known exactly.\cite{Rokhsar:1988ko}
This state forms a disordered liquid like state which is, depending on the lattice, either gapless or gapped.
In particular, it has been shown that the ground state for the \textsc{QDM} on the square lattice at the \textsc{RK}~point is a gapless $\mathrm{U}(1)$~liquid, while the same model on the triangular lattice forms a gapped topologically ordered $\mathbb{Z}_2$~liquid.\cite{Moessner2001}
The topological order found in the \textsc{QDM} on the triangular lattice is the same as the one found in the toric code (\textsc{TC}) model (i.e. the two models have the same quasi-particle excitations and thus the same $U$- and $S$-matrices).\cite{Kitaev2003}
Misguich et al. derived a different type of \textsc{QDM} on the kagome lattice which consists of commuting terms and thus can be solved exactly for the ground state and all excited states.\cite{Misguich:2002fa}
This \textsc{QDM} can be exactly mapped to the \textsc{TC}~model on the honeycomb lattice.

In this paper, we demonstrate how to construct an exactly solvable \textsc{QDM} that realizes a different type of topological order, namely the so-called double semion (\textsc{DSem}) phase first proposed in Ref.~\onlinecite{Freedman2004}.
It was later generalized to \enquote{string-net} models which realize various other kinds of topological order via condensation of extended objects.\cite{Levin2005}
The \textsc{DSem}~phase can also be understood as a $\mathbb{Z}_2$~gauge theory twisted by a nontrivial 3-cocycle.\cite{Dijkgraaf1990,Buerschaper:2013ud}
The main idea of our approach is to convert the known loop-gas representation of the \textsc{DSem}~phase\cite{Freedman2004} on the honeycomb lattice into a \textsc{QDM} on the kagome lattice.
Starting from the exactly solvable point, we can prove that a stable \textsc{DSem}~phase is formed by numerically obtaining the braiding statistics  using exact diagonalization of small clusters.
Our model thus presents an approach to realize more exotic types of topological order in frustrated spin systems.

This paper is organized as follows.
We start by deriving the exactly solvable \textsc{QDM} in Sec.~\ref{sec:exact} and show that its ground state has \textsc{DSem} topological order.
Next, we explore the stability of the phase in Sec.~\ref{sec:stability} by perturbing the model away from its exactly solvable point.
We then examine possible realizations of the \textsc{DSem}~phase in a frustrated spin system in Sec.~\ref{sec:realization}.
We finally conclude with a brief summary and outlook in Sec.~\ref{sec:conclusion}.

\section{Exactly Solvable Models}
\label{sec:exact}

First we briefly review the \textsc{TC} and \textsc{DSem}~models on the honeycomb lattice which realize the two distinct types of $\mathbb{Z}_2$~topological order.\cite{Kitaev2003,Freedman2004}
We then construct two corresponding, exactly solvable \textsc{QDM}s by mapping the charge-free subspace of spins on the honeycomb lattice to dimer coverings of the kagome lattice via an intermediate arrow representation.
The first \textsc{QDM} coincides exactly with the one derived earlier.\cite{Misguich:2002fa}
The second \textsc{QDM} is new and realizes the type of $\mathbb{Z}_2$~topological order found in the \textsc{DSem}~phase.

\subsection{Toric Code and Double Semion Models on the Honeycomb Lattice}

\begin{figure}
    \includegraphics[width=\columnwidth]{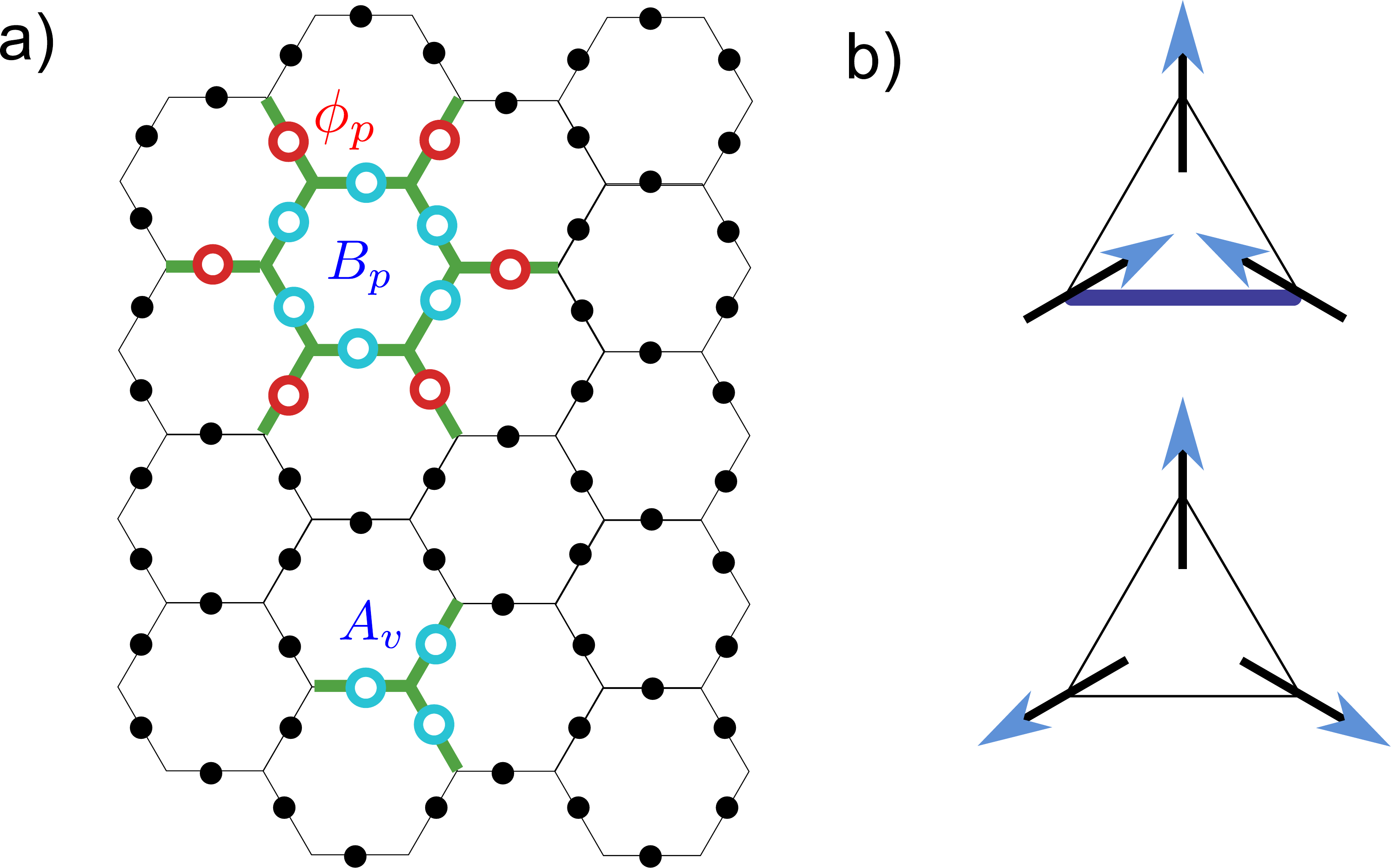}
    \caption{\label{fig:loop_space}
        a) Terms in the Hamiltonians~\eqref{eq:Hamiltonians} on the honeycomb lattice.
        b) Mapping from arrows to dimers.
        Arrows live on edges of the honeycomb lattice and hence the sites of the kagome lattice.}
\end{figure}

\begin{figure*}
    \includegraphics[scale=0.24]{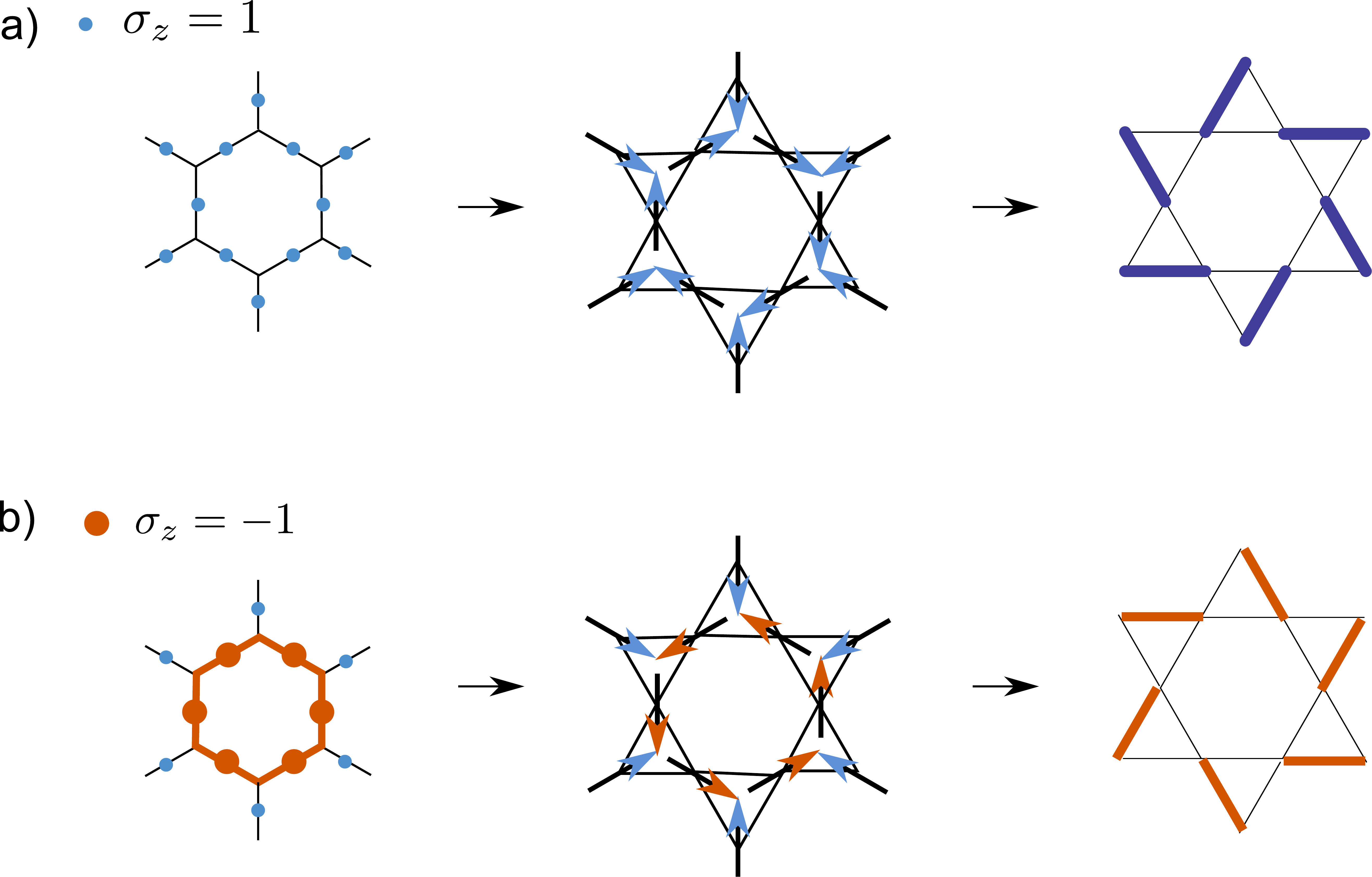}
    \caption{\label{fig:arrow}
        a) Mapping from spin space on the honeycomb lattice to a reference arrow configuration and finally to a corresponding reference dimer covering on the kagome lattice.
        b) Flipping spins on the honeycomb lattice corresponds to flipping all inner arrows around a hexagon and hence to a local dimer resonance.}
\end{figure*}

The models are defined by the following Hamiltonians which acts on spins ($\sigma^z=\pm 1$) living on the edges (Fig.~\ref{fig:loop_space}a) of a honeycomb lattice
\begin{equation}
    \label{eq:Hamiltonians}
    H^{\mathrm{TC},
       \mathrm{DSem}}
    =-
     \sum_v
     A_v-
     \sum_p
     B_p
     \phi_p^{\mathrm{TC},
             \mathrm{DSem}},
\end{equation}
where for both models
\begin{equation}
    A_v
    =\smashoperator[l]{\prod_{k
                              \in
                               E(v)}}\mkern-4mu
     \sigma_k^z,
    \qquad
    B_p
    =\smashoperator[l]{\prod_{k
                              \in
                               E(p)}}
     \sigma_k^x.
\end{equation}
and the phase factors~$\phi_p$ are given by 
\begin{equation}
    \phi_p^\mathrm{TC}
    =I,
    \qquad
    \phi_p^\mathrm{DSem}
    =-
     \Bigl[\prod_{j
                  \in
                   \tilde{E}(p)}
           \mathrm{i}^{(1-
                        \sigma_j^z)/
                       2}\Bigr]\mkern2mu
     \mathcal{P}_v.
\end{equation}
Here $E(v)$ is the set of edges around a vertex~$v$, $E(p)$ the set of inner edges  and $\tilde{E} (p)$ is the set of outer edges around a hexagon~$p$.
The common $+1$~eigenspace of all vertex terms~$A_v$ is usually called the \emph{charge-free} subspace~$\mathcal{L}_\mathrm{S}$.
If we imagine an edge of the honeycomb lattice in spin state~$\ket{1}$ to represent a short string then a basis of~$\mathcal{L}_\mathrm{S}$ is given by spin configurations whose strings form closed loops.
Conversely, a spin configuration with some open string lies in~$\mathcal{L}_\mathrm{S}^\perp$.
We denote the projector onto~$\mathcal{L}_\mathrm{S}$ by~$\mathcal{P}_v$.

We can easily see that the plaquette terms ($B_p \phi_p^\mathrm{TC}$, $B_p \phi_p^\mathrm{DSem}$) and the vertex term ($A_v$) commute since the plaquette terms only flip a pair of spins at each vertex.
As sums of commuting terms the Hamiltonians are thus exactly solvable.
On a sphere, the ground state of the \textsc{TC}~model is the common $+1$~eigenstate of all $B_p \phi_p^\mathrm{TC}$ and $A_v$ terms and reads
\begin{equation}
    \ket{\psi}
    =\sum_C\mkern2mu
     \ket{C},
\end{equation}
where $C$ is a configuration with closed loops.
In other words, the ground state is an equal weight superposition of closed loop configurations.
Similarly, the ground state of the \textsc{DSem}~model is the common $+1$ eigenstate of all $B_p \phi_p^\mathrm{DSem}$ and $A_v$ terms and is given by
\begin{equation}
    \ket{\psi}
    =\sum_C
     (-1)^{N_C}
     \ket{C},
\end{equation}
where $N_C$ is the number of loops in a configuration~$C$ with closed loops.

When placed on a torus (assumed henceforth), both models show a four-fold degeneracy and the degenerate ground states can be classified into different topological sectors distinguished by the number of large loops around the two directions of the torus modulo two. 
There are four anyonic excitations in each system, but their statistics are fundamentally different.
The excitations in the \textsc{TC}~model consist of three particles with trivial self-statistics (identity particle, charge 1 electric defect, $\pi$-flux magnetic defect) and a fermion (bound state of charge and flux). 
In contrast, the \textsc{DSem}~model has two semions of opposite chirality and two bosons (identity particle and a $\pi$-flux defect which is a bound state of the semions).
We note that the excitations are deconfined and can only be created in pairs.
For a more detailed description see Refs.~\onlinecite{Kitaev2003,Freedman2004,Levin2005}.

\subsection{Intermediate Arrow Representation}
\label{sec:arrows}

In the following we will assign arrows to the edges of the honeycomb lattice.
This arrow representation was introduced earlier in Refs.~\onlinecite{Elser:1993ja,Misguich:2002fa}, together with an additional vertex rule (explained below).
Here we will relax this vertex rule in order to represent the complete Hilbert space of spins by arrows.

An arrow configuration is said to obey the vertex rule at vertex~$v$ if the number of incoming arrows at~$v$ is even.
If an arrow configuration obeys the vertex rule everywhere we say it belongs to the charge-free arrow configurations~$\mathcal{L}_\mathrm{A}$, and to~$\mathcal{L}_\mathrm{A}^\perp$ otherwise.
Now fix a random arrow configuration~$D_0$ in~$\mathcal{L}_\mathrm{A}$.
Given some arrow configuration~$D$, each arrow represents a local spin in an eigenstate of~$\sigma^z$.
We define the arrow to represent~$\ket{0}$ if it is aligned with the corresponding arrow in~$D_0$, and~$\ket{1}$ otherwise.
This clearly defines a local one-to-one identification of the complete Hilbert spaces of arrows and spins.
The reference arrow configuration~$D_0$ itself represents the polarized spin state~$\ket{0\dots 0}$.
The action of~$\sigma^x$ is defined by a flip in the arrow representation and thus translates to the usual $\ket{i}\mapsto\ket{i\oplus 1}$ in the spin representation, for any~$i\in\set{0,1}$.

If we choose a different reference arrow configuration~$D_0'$ in~$\mathcal{L}_\mathrm{A}$ our original definition of~$\sigma^z$ changes to $-\sigma^z$ for every lattice edge whose arrow in~$D_0'$ is not aligned with the one in~$D_0$.
Since the definition of~$\sigma^x$ is \emph{not} affected we can write this as $\sigma^z\mapsto\sigma^x\sigma^z\sigma^x$ locally.
Thus changing the reference arrow configuration corresponds to a local unitary circuit~$\mathcal{U}$ of depth~$1$.
Note that any such~$\mathcal{U}$ consists of loops of~$\sigma^x$ on the lattice.

With this definition in place, the Hamiltonians~\eqref{eq:Hamiltonians} are equally valid for both the spin and the arrow representation.
While their action on spins is clear, let us briefly sketch how they act on an arrow configuration~$D$.
The term~$A_v$ yields an eigenvalue~$+1$ whenever $D$ obeys the vertex rule at~$v$, and~$-1$ otherwise.%
\footnote{Indeed, $D_0$ obeys the vertex rule and represents $\ket{000}$ by definition.
    An arrow configuration~$D$ obeys the vertex rule iff it can be obtained from~$D_0$ by flipping an even number of arrows.}
This means that the definition of the arrow representation identifies the subspaces~$\mathcal{L}_\mathrm{A}$ and~$\mathcal{L}_\mathrm{S}$.
The term~$B_p$ flips all (inner) arrows around a hexagon~$p$.
In the \textsc{DSem} model the term~$-\phi_p$ adds an additional phase factor~$\mathrm{i}$ for each outer arrow in~$D$ which is not aligned with the one in~$D_0$.
If $D$ is charge-free it is easy to see that the number of these misaligned arrows is always even, so the action of~$H$ on charge-free arrow configurations is Hermitian.
Clearly, the ground states of both models on a sphere are superpositions of all charge-free arrow configurations with weights~$\pm 1$.

If we choose a different reference arrow configuration~$D_0'$ our Hamiltonians~\eqref{eq:Hamiltonians} change to $H'=\mathcal{U}H\mathcal{U}^\dagger$, where $\mathcal{U}$ is the above local unitary circuit.
It is immediate that $\mathcal{U}$ commutes with all~$A_v$ and~$B_p$, hence the Hamiltonian of the \textsc{TC}~model is \emph{independent} of the choice of reference arrow configuration.
For the \textsc{DSem}~model on the other hand the Hamiltonian \emph{depends} on this choice, namely through $\phi_p'=\mathcal{U}\phi_p \mathcal{U}^\dagger$ which introduces the nontrivial phase factors~$\pm 1$.
This is actually not too surprising: while $\mathcal{U}$ respects the decomposition $\mathcal{L}_\mathrm{A}\oplus\mathcal{L}_\mathrm{A}^\perp$ of the arrow Hilbert space it may permute charge-free arrow configurations arbitrarily.

\subsection{Dimer Representation and \textsc{QDM}s}

Finally we study dimer coverings of the kagome lattice which is obtained as the medial lattice of the honeycomb lattice considered so far.
A basis of this Hilbert space~$\mathcal{L}_\mathrm{D}$ is given by dimer coverings with the property that there is exactly one hard-core dimer around every vertex.
In contrast to \textsc{RVB}~states, these basis states are assumed to be orthogonal.

Now we map \emph{charge-free} arrow configurations on the honeycomb lattice to these dimer coverings in the usual way~\cite{Elser:1993ja,Misguich:2002fa} (see Fig.~\ref{fig:loop_space}b).
Note that this induces a bijection between charge-free \emph{spin} configurations~$\mathcal{L}_\mathrm{S}$ on the honeycomb lattice and dimer coverings~$\mathcal{L}_\mathrm{D}$ of the kagome lattice.
This mapping \emph{depends} on the choice of a reference dimer covering~$D_0$ (which we identify with the reference arrow configuration above).

If we restrict the Hilbert space to charge-free states the Hamiltonians of the \textsc{TC} and \textsc{DSem}~model have a natural action on dimer coverings.
It is instructive to study this in more detail.
In both cases the arrow flips of the term~$B_p$ translate into 32~dimer resonance moves around the hexagon~$p$.\cite{Zeng:1995jv}
Each such dimer resonance move corresponds to a loop~$\alpha$ around~$p$ with $\abs{\alpha}=2n$ edges ($3\leq n\leq6$), and $d_\alpha(p)$ and $\bar{d}_\alpha(p)$ denote the two ways in which $n$~dimers can be placed along that loop.\cite{Misguich:2002fa}
In the \textsc{DSem}~model the term~$\phi_p$ adds an additional phase factor $f_\alpha(p)=\pm 1$ to each dimer resonance move.
We can now rewrite~\eqref{eq:Hamiltonians} as
\begin{equation}
    \label{eq:Hamiltonian_QDM}
    H^{\mathrm{TC},
       \mathrm{DSem}}
    =\sum_p
     \sum_{n
           =3}^6
     h_n(p),
\end{equation}
where
\begin{equation}
    \label{eq:h_def}
    h_n(p)
    =\smashoperator[l]{\sum_{\abs{\alpha}
                             =2
                              n}}
     f_\alpha(p)\mkern2mu
     \bigl(\ketbra{d_\alpha(p)}
                  {\bar{d}_\alpha(p)}+
           \ketbra{\bar{d}_\alpha(p)}
                  {d_\alpha(p)}\bigr)
\end{equation}
collects all resonance moves involving the same number of dimers.

\begin{table}
    \begin{ruledtabular}
        \begin{tabular}{cccccc}
            $n$ &                            $\alpha$                           & $f_\alpha(p_0)$ & $f_\alpha(p_1)$ & $f_\alpha(p_2)$ & $f_\alpha(p_3)$ \\
            \colrule
            3 & \vcorrect{12bp}{\includegraphics[scale=0.3]{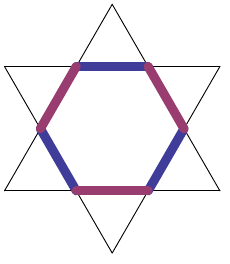}}    &       $+1$      &       $+1$      &       $+1$      &       $+1$      \\
            \colrule
            4 & \vcorrect{12bp}{\includegraphics[scale=0.3]{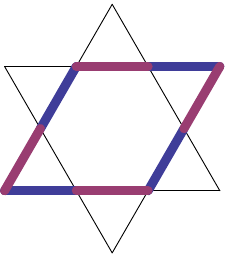}}
                \vcorrect{12bp}{\includegraphics[scale=0.3]{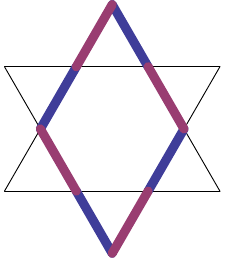}}
                \vcorrect{12bp}{\includegraphics[scale=0.3]{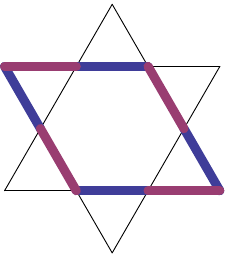}}  &       $-1$      &       $-1$      &       $-1$      &       $-1$      \\
            \colrule
            4 & \vcorrect{12bp}{\includegraphics[scale=0.3]{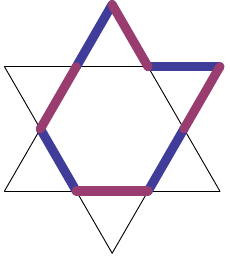}}
                \vcorrect{12bp}{\includegraphics[scale=0.3]{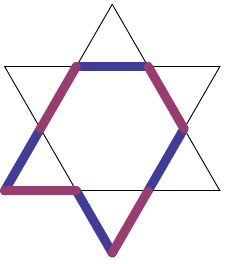}} &       $-1$      &       $-1$      &       $+1$      &       $+1$      \\
              & \vcorrect{12bp}{\includegraphics[scale=0.3]{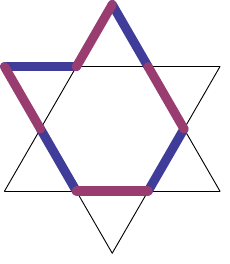}}
                \vcorrect{12bp}{\includegraphics[scale=0.3]{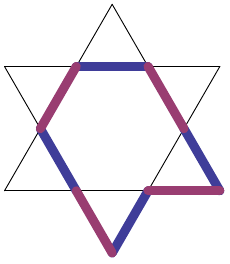}} &       $-1$      &       $+1$      &       $-1$      &       $+1$      \\
              & \vcorrect{12bp}{\includegraphics[scale=0.3]{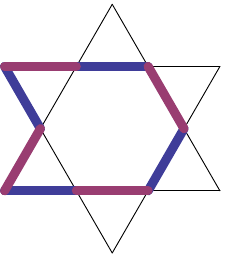}}
                \vcorrect{12bp}{\includegraphics[scale=0.3]{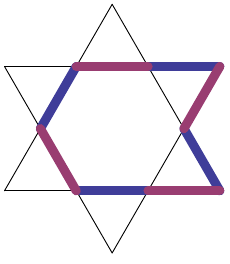}} &       $-1$      &       $+1$      &       $+1$      &       $-1$      \\
            \colrule
            4 & \vcorrect{12bp}{\includegraphics[scale=0.3]{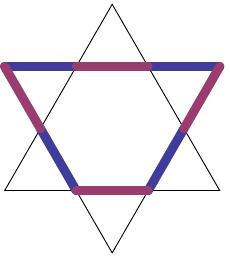}}
                \vcorrect{12bp}{\includegraphics[scale=0.3]{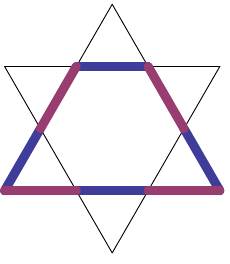}} &       $-1$      &       $+1$      &       $+1$      &       $-1$      \\
              & \vcorrect{12bp}{\includegraphics[scale=0.3]{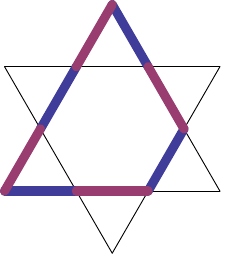}}
                \vcorrect{12bp}{\includegraphics[scale=0.3]{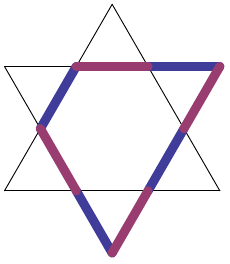}} &       $-1$      &       $-1$      &       $+1$      &       $+1$      \\
              & \vcorrect{12bp}{\includegraphics[scale=0.3]{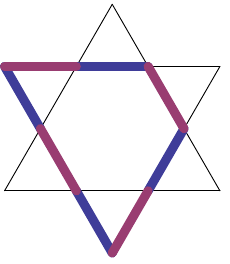}}
                \vcorrect{12bp}{\includegraphics[scale=0.3]{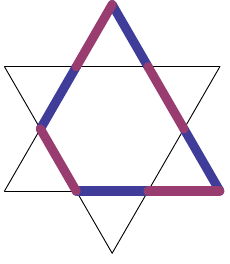}} &       $-1$      &       $+1$      &       $-1$      &       $+1$      \\
            \colrule
            5 & \vcorrect{12bp}{\includegraphics[scale=0.3]{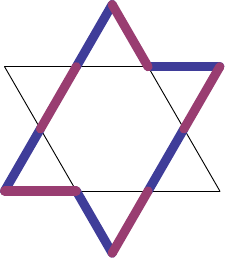}}
                \vcorrect{12bp}{\includegraphics[scale=0.3]{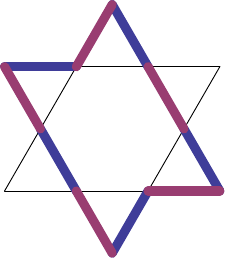}}
                \vcorrect{12bp}{\includegraphics[scale=0.3]{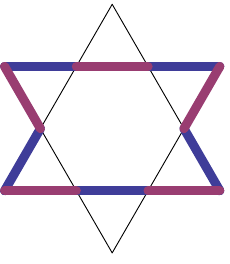}}  &       $+1$      &       $+1$      &       $+1$      &       $+1$      \\
            \colrule
            5 & \vcorrect{12bp}{\includegraphics[scale=0.3]{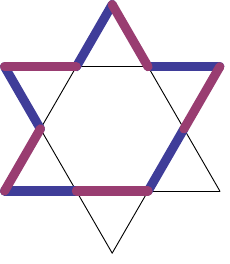}}
                \vcorrect{12bp}{\includegraphics[scale=0.3]{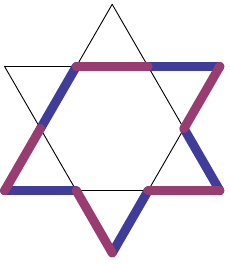}} &       $+1$      &       $-1$      &       $+1$      &       $-1$      \\
              & \vcorrect{12bp}{\includegraphics[scale=0.3]{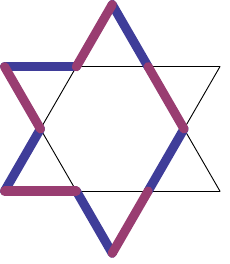}}
                \vcorrect{12bp}{\includegraphics[scale=0.3]{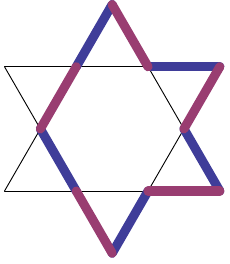}} &       $+1$      &       $-1$      &       $-1$      &       $+1$      \\
              & \vcorrect{12bp}{\includegraphics[scale=0.3]{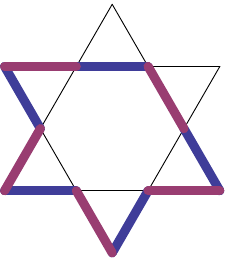}}
                \vcorrect{12bp}{\includegraphics[scale=0.3]{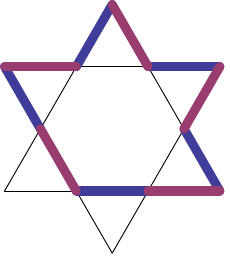}} &       $+1$      &       $+1$      &       $-1$      &       $-1$      \\
            \colrule
            5 & \vcorrect{12bp}{\includegraphics[scale=0.3]{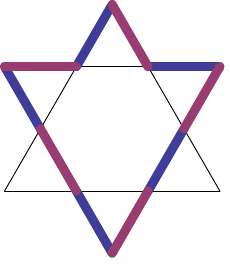}}
                \vcorrect{12bp}{\includegraphics[scale=0.3]{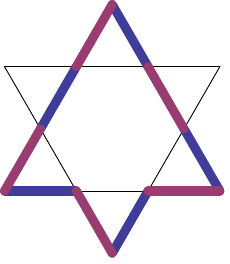}} &       $+1$      &       $-1$      &       $-1$      &       $+1$      \\
              & \vcorrect{12bp}{\includegraphics[scale=0.3]{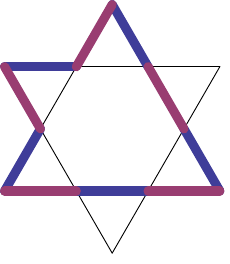}}
                \vcorrect{12bp}{\includegraphics[scale=0.3]{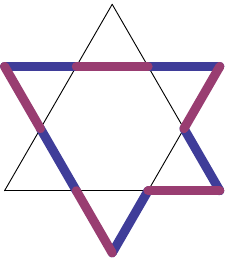}} &       $+1$      &       $+1$      &       $-1$      &       $-1$      \\
              & \vcorrect{12bp}{\includegraphics[scale=0.3]{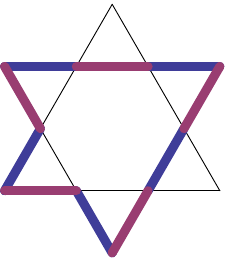}}
                \vcorrect{12bp}{\includegraphics[scale=0.3]{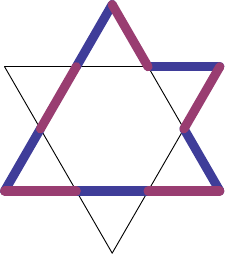}} &       $+1$      &       $-1$      &       $+1$      &       $-1$      \\
            \colrule
            6 & \vcorrect{12bp}{\includegraphics[scale=0.3]{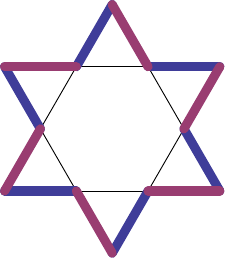}}    &       $-1$      &       $-1$      &       $-1$      &       $-1$      \\
        \end{tabular}
    \end{ruledtabular}
    \caption{\label{tab:resonances}
        Dimer resonance moves of the canonical Hamiltonian realizing the \textsc{DSem}~universality class.
        For each loop~$\alpha$ the resonant dimer coverings~$d_\alpha(p)$ and~$\bar{d}_\alpha(p)$ are shown in red and blue respectively.
        The phase factors~$f_\alpha(p)$ depend on the hexagons~$p_0$, …, $p_3$ which form an \emph{enlarged} unit cell on the kagome lattice (see Fig.~\ref{fig:ED_72}b).}
\end{table}

\begin{figure}
    \includegraphics[width=\columnwidth]{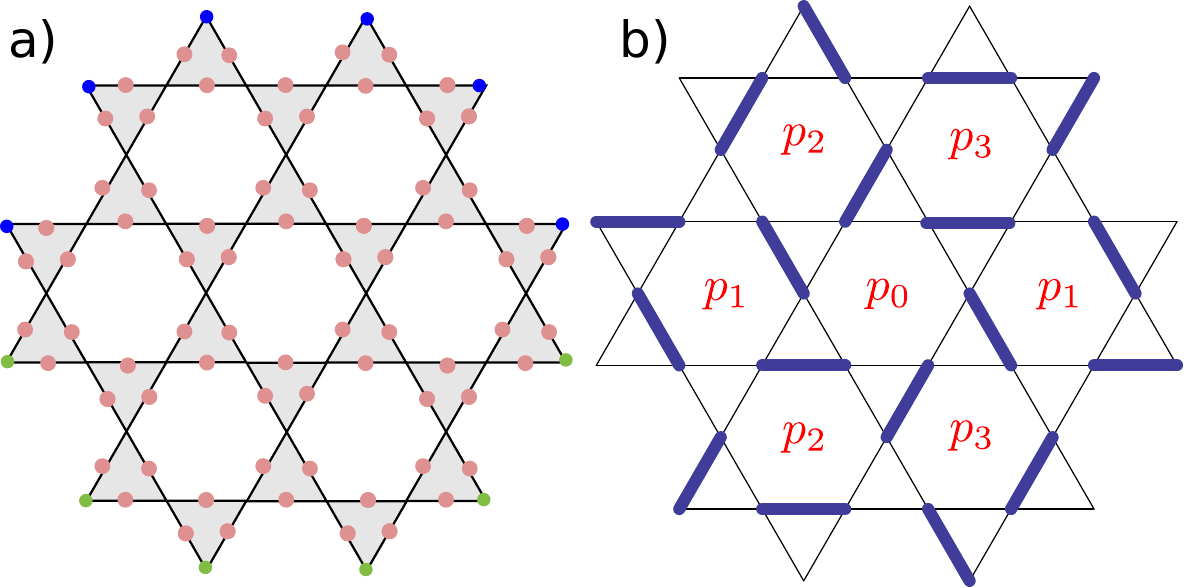}
    \caption{\label{fig:ED_72}
       a) 72-site cluster on a torus used in the exact diagonalization study.
       Red circles indicate dimers while blue and green circles on opposite sides of the cluster indicate dimers identified by periodic boundary conditions.
       b) Reference dimer covering with a $2\pi/3$-rotation symmetry.}
\end{figure}

For the \textsc{TC}~model, all phase factors~$f_\alpha(p)$ are trivial and \eqref{eq:Hamiltonian_QDM} reduces to the \textsc{QDM} Hamiltonian in Ref.~\onlinecite{Misguich:2002fa}.
For the \textsc{DSem}~model, the phase factors~$f_\alpha(p)$ \emph{depend} on the reference dimer covering~$D_0$ and it seems that this may influence the form of the Hamiltonian dramatically.
For example, a random~$D_0$ typically leads to a Hamiltonian without any lattice symmetries.
Yet, we showed in Sec.~\ref{sec:arrows} that any two of these Hamiltonians are equivalent to each other up to a local unitary circuit~$\mathcal{U}$ of depth~1.
Since $\mathcal{U}$ cannot change the universality class of the models we can define a particularly simple, \emph{canonical} Hamiltonian for the \textsc{DSem}~universality class by choosing a reference dimer covering~$D_0$ which is invariant under rotations by~$2\pi/3$ and translations by \emph{two} unit cells of the kagome lattice (see Fig.~\ref{fig:ED_72}b).
The resulting phase factors~$f_\alpha(p)$ are then given in Tab.~\ref{tab:resonances}.

Like the corresponding models in the spin space, these models in the dimer space are also exactly solvable.
Their ground states have a correlation length of exactly one lattice spacing and show topological order.
We can also construct their anyonic excitations by acting with string operators as indicated in Ref.~\onlinecite{Keyserlingk2013} (appropriately redefined in the dimer space).

\section{Stability of the Phase}
\label{sec:stability}

\begin{figure}
    \includegraphics[width=\columnwidth]{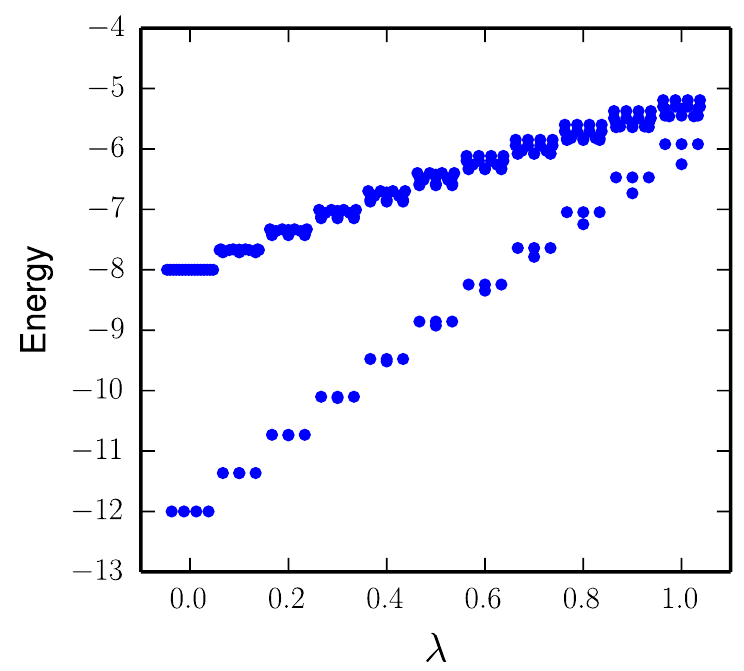}
    \caption{\label{fig:gap_stability}
	    Energy spectrum of the Hamiltonian~\eqref{eq:Hamiltonian_interpolation}.
	    We observe a topologically degenerate ground state sector with a finite gap to all excitations.
	    This gap does not close when tuning the system from the exactly solvable point to $H_4$.}
\end{figure}

We now explore the stability of the \textsc{DSem} dimer model defined in~\eqref{eq:Hamiltonian_QDM}.
Clearly, the term~$h_3$ only has trivial phase factors and is not ergodic in the space of dimer coverings.
In contrast, the term~$h_4$ exhibits a rich structure of non-trivial phase factors as seen in Tab.~\ref{tab:resonances}. 
We also note that the $h_4$~term is ergodic in the dimer space.
This ergodicity has already been seen in the \textsc{TC} dimer model.\cite{Melko2014}
A very similar argument can be made for the \textsc{DSem} dimer model and this suggests that the $h_4$~term is enough to realize the \textsc{DSem}~phase.
To verify this, we interpolate between the exactly solvable model~\eqref{eq:Hamiltonian_QDM} and~$h_4$ via
\begin{equation}
    \label{eq:Hamiltonian_interpolation}
    H(\lambda)
    =\lambda
     H^\mathrm{DSem}+
     (1-
      \lambda)
     H_4,
\end{equation}
where $H_4=\sum_p h_4(p)$.
We consider the above model on a 72-site cluster placed on a torus (see Fig.~\ref{fig:ED_72}a).
We use exact diagonalization (\textsc{ED}) in the space of dimer coverings and obtain four topologically degenerate ground states and indicate the spectrum in Fig.~\ref{fig:gap_stability}.
The ground state gap obtained from \textsc{ED} does not seem to close indicating that we do not pass any phase transition during the interpolation.
This suggests that both Hamiltonians are in the same phase. 
To confirm this, we obtain the modular $U$- and $S$-matrices characterizing the braiding statistics of the quasi-particle excitations.

The $U$- and $S$-matrices encode the exchange statistics and mutual statistics of the excitations respectively.
In order to obtain them, we first find the minimally entangled states (\textsc{MES}) of the system.
These are ground states which minimize the entanglement on non-trivial bipartitions of the torus.
We then obtain our matrices from the action of a $2\pi/3$-rotation on the \textsc{MES} $\set{\ket{\Xi_i}}$:\cite{Zhang2012,Cincio2013}
\begin{equation}
    (D^\dagger
     U
     S
     D)_{i
         j}
    =\expval{\Xi_j}
            {R_{2
                \pi/
                3}}
            {\Xi_i},
\end{equation}
where $D$ is a diagonal matrix containing the arbitrary phases that can come with each~$\ket{\Xi_i}$.
We then obtain the $U$- and $S$-matrices individually from the matrix~$US$.\cite{Morampudi2014}
We find the following modular matrices for~$H_4$:
\begin{align}
    U^{H_4}
    & =U^\mathrm{DSem}+
       10^{-1}
       \begin{pmatrix}
           3.0 &     &                                           &                      0                     \\
               & 0.9 &                                           &                                            \\
               &     & 1.2\mkern2mu\mathrm{e}^{0.4\mathrm{i}\pi} &                                            \\
            0  &     &                                           & 1.0\mkern2mu\mathrm{e}^{-0.2\mathrm{i}\pi}
       \end{pmatrix}, \\
    S^{H_4}
    & =S^\mathrm{DSem}+{}
       \nonumber \\
    & \hphantom{={}}
       10^{-1}
       \begin{pmatrix}
           -0.6 &                    -0.6                   &                    -0.6                    &                    -0.6                    \\
           -0.6 &                     0.2                   & 0.4\mkern2mu\mathrm{e}^{-0.3\mathrm{i}\pi} & 0.4\mkern2mu\mathrm{e}^{-0.3\mathrm{i}\pi} \\
           -0.6 & 0.1\mkern2mu\mathrm{e}^{0.9\mathrm{i}\pi} &  0.7\mkern2mu\mathrm{e}^{0.9\mathrm{i}\pi} &                     0.7                    \\
           -0.6 &                     0.5                   &  0.5\mkern2mu\mathrm{e}^{0.3\mathrm{i}\pi} & 0.6\mkern2mu\mathrm{e}^{-0.7\mathrm{i}\pi}
       \end{pmatrix}.
\end{align}
This again strongly indicates that $H_4$ is in the \textsc{DSem}~phase even though the model is not exactly solvable any more.

\section{Spin Model}
\label{sec:realization}


\begin{figure}
    \includegraphics[width=\columnwidth]{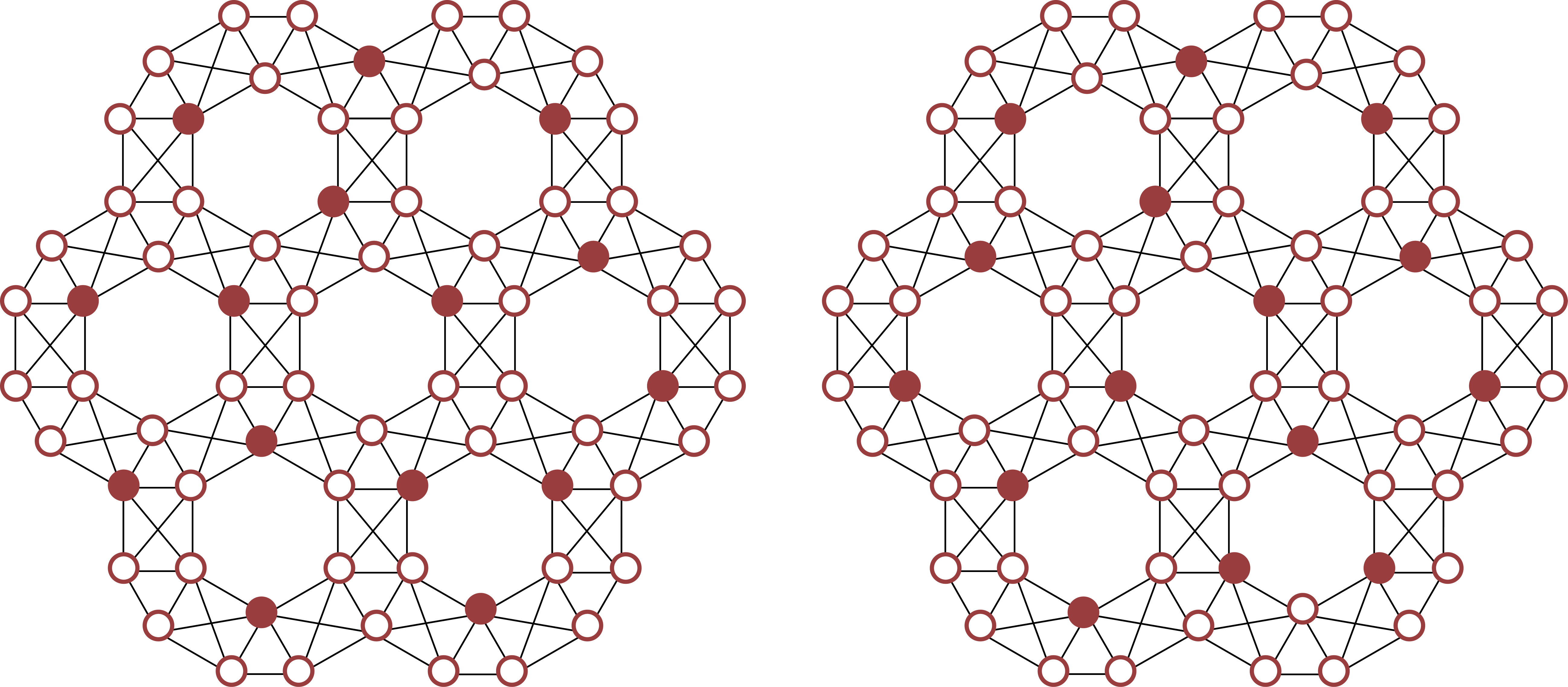}
    \caption{\label{fig:ruby}
        Spins lives on the sites of a modified ruby lattice obtained as the medial lattice of the kagome. Filled circles denote up-spins and empty circles denote down-spins. Two of many degenerate configurations are shown here. }
\end{figure}

While the quantum dimer model discussed above is usually motivated by resonating-valence bond states (which are potentially realized in the spin-1/2 Heisenberg model on the kagome lattice~\cite{Misguich:2003bd}), we introduce here a different route. 
In particular, we consider an XXZ model on a modified \emph{ruby}  lattice~\cite{Hu2011,Wu2012} in the strongly anisotropic limit (see Fig.~\ref{fig:ruby}). 
For this model, we show that the low-energy physics is well described by the TC version of our quantum dimer model.
The XXZ Hamiltonian is given by

\begin{equation}
\label{eq:XXZ}
H = -\frac{J_{\bot}}{2} \sum_{\langle i, j \rangle} (S^{+}_{i} S^{-}_{j} + h.c. ) + J_z \sum_{\langle i, j \rangle}  S^{z}_{i}S^{z}_{j} - h \sum_i S^{z}_{i},
\end{equation}

\noindent where $\sum_{\langle i, j \rangle}$ runs over nearest-neighboring sites of the modified ruby lattice, $J_z,J_{\bot}\ge0$ are the coupling parameters, and the operators $S_i^{+/-},S_i^z$  are the usual spin-1/2 operators. 
Clearly, the Hamiltonian conserves the total magnetization $m=\sum_iS_i^z$.
In the following, we study the limit  of $J_{\bot} \ll J_z$ and consider the case of 1/4 magnetization (which can be reached by tuning the external magnetic field $h$).
We find an extensive ground state degeneracy for $J_{\bot} = 0$: \emph{All configurations that have exactly three up-spins and one down-spin per crossed square are ground states}.
As the sites of the ruby lattice correspond to the bonds of the kagome lattice, the ground state manifold  maps exactly on the dimer manifold. 
The degeneracy is then lifted by quantum fluctuations for any finite  $J_{\bot}$. 
In the limit $J_{\bot} \ll J_z$,  we obtain perturbatively an effective  model acting on the dimer manifold,   
\begin{equation}
\label{eq:e_QDM}
H_{D} = 12\frac{J^{3}_{\bot}}{J^{2}_z} H_{3} + 44\frac{J^{4}_{\bot}}{J^{3}_z}H_{4}
\end{equation}
\noindent up to 4th order in perturbation theory.
Here, $H_i = \sum_p h_i (p)$ with $h_i (p)$ defined in Eq.~(\ref{eq:h_def})  with  phase factors $f_{\alpha} (p) = 1$ for all $p$.
In the previous section, we observed that the $H_{4}$ term was sufficient to realize the topologically ordered phase.
Thus we might expect that the effective Hamiltonian might stabilize a TC phase for some parameter regime.
By exact diagonalization of Hamiltonian Eq.~(\ref{eq:e_QDM}), we first calculate the topological entanglement entropy\cite{Kitaev2006, Levin2006} (TEE) of the two models on the 72 site cluster as a function of $J_{\bot}/J_z$.
The TEE is a length-independent, universal correction $\gamma$ to the area law for the entanglement entropy $S = \alpha L - \gamma$, where $L$ is the length of the boundary of the subsystem and $\alpha$ a non-universal constant.
As the TC phase is an abelian phase with four quasiparticles, it is characterized by $\gamma = \log 2$. 
For $J_{\bot}/J_z \lesssim 0.2$, we find a unique ground state and $\gamma\approx 0$. 
This is because the $H_3$ term dominates and creates a trivial ground state in which the spins resonate around hexagons.
For $J_{\bot}/J_z > 0.2$, we find four ground states that are approximately degenerate and $\gamma\approx 0.5$, indicating the presence of a topological phase.
We attribute the deviation from $\log 2$ to the strong finite size which prevent us from attaining the exact value on this small cluster.
Additionally, we calculate the modular matrices which turn out to be a more robust indicator of the topological phase. For the parameter $J_{\bot}/J_{z}=0.3$ we find that
\begin{align}
    U^{0.3}
    & =U^\mathrm{TC}+
       10^{-1}
       \begin{pmatrix}
           2.5 &     &                                           &                      0                     \\
               & 0.3 &                                           &                                            \\
               &     & 1.0\mkern2mu\mathrm{e}^{0.4\mathrm{i}\pi} &                                            \\
            0  &     &                                           & 0.9\mkern2mu\mathrm{e}^{-0.1\mathrm{i}\pi} \\
       \end{pmatrix}, \\
    S^{0.3}
    & =S^\mathrm{TC}+{}
       \nonumber \\
    & \hphantom{={}}
       10^{-1}
       \begin{pmatrix}
           -0.5 & -0.5 & -0.5 & -0.5 \\
           -0.5 &  0.4 & 0.3\mkern2mu\mathrm{e}^{-0.4\mathrm{i}\pi} & 0.3\mkern2mu\mathrm{e}^{-0.4\mathrm{i}\pi} \\
           -0.5 & 0.3\mkern2mu\mathrm{e}^{0.9\mathrm{i}\pi} & 0.5\mkern2mu\mathrm{e}^{0.9\mathrm{i}\pi} & 0.5\mkern2mu\mathrm{e}^{-0.1\mathrm{i}\pi} \\
           -0.5 & 0.4\mkern2mu\mathrm{e}^{0.1\mathrm{i}\pi} & 0.4\mkern2mu\mathrm{e}^{0.4\mathrm{i}\pi} & 0.5\mkern2mu\mathrm{e}^{-0.6\mathrm{i}\pi}
       \end{pmatrix}.
\end{align}
The $U$ and $S$ matrices obtained  correspond to the TC~topological order. 
This provides  numerical indications that the considered XXZ model does exhibit a TC code phase for a range of parameters in the limit of  $J_{\bot} \ll J_z$. 
We note that by choosing particular phases in the off-diagonal part of Hamiltonian Eq.~(\ref{eq:XXZ}), it is possible to also realize the \textsc{DSem}~phase in a similar manner. 
Though the Hamiltonian will be considerably more complex, there is a hope that it might be realized in a cold atom setting. 
This and the study of the full XXZ model will be subject to a future work.
\section{Conclusions}
\label{sec:conclusion}

In this paper, we constructed an exactly solvable \textsc{QDM} on the kagome lattice which realizes the \textsc{DSem}~phase.
Our derivation employs local unitary circuits of constant depth to identify the spin space of the \textsc{DSem}~model on the honeycomb lattice with the space of dimer coverings of the kagome lattice via an intermediate arrow representation.
While the \textsc{QDM} Hamiltonian depends on the choice of a reference dimer covering, we showed that we can always restrict to a canonical Hamiltonian which respects a maximal number of symmetries of the kagome lattice, namely, rotations by~$2\pi/3$ and translations by \emph{two} unit cells.
This weak breaking of translation symmetry, which appears inevitably in our construction, may very well turn out to be a general feature of spin models in the \textsc{DSem}~phase.
We would also like to emphasize that not only the ground states but in fact all excitations of our \textsc{QDM} can be obtained exactly.
Our results can thus be viewed as a natural generalization of Ref.~\onlinecite{Misguich:2002fa} to a unified implementation of the two distinct types of $\mathbb{Z}_2$ topological order within exactly solvable \textsc{QDM}s. 

Furthermore, we explored the stability of our \textsc{QDM} using numerical exact diagonalization.
We established that a considerably reduced \textsc{QDM} away from the exactly solvable point still lies in the \textsc{DSem}~phase as evidenced by its modular $U$- and $S$-matrices which characterize the statistics of the excitations.
We finally constructed a spin model which realizes the \textsc{TC}~phase and indicated possible extensions to also realize the \textsc{DSem}~phase.

This work can be immediately generalized to all lattices consisting of corner-sharing triangles and beyond, for example, to the star lattice%
\footnote{We start from a honeycomb lattice and clone each edge into two new ones~$i$ and~$j$.
    For the spin representation this adds an additional term $\sigma_i^z\sigma_j^z$ per original edge.
    This local procedure clearly does not change the universality class of the \textsc{DSem}~model, but allows as to exploit an arrow representation.}.
Other extensions of this work include possible generalizations to other string-net models, in particular to those realizing non-Abelian phases.
This could possibly lead to concrete realizations of proposals such as universal quantum computation.


\begin{acknowledgments}
    We thank Roderich Moessner, Cécile Repellin and in particular Kirill Shtengel for useful discussions.
    This research was supported in part by Perimeter Institute for Theoretical Physics.
    Research at Perimeter Institute is supported by the Government of Canada through Industry Canada and by the Province of Ontario through the Ministry of Research and Innovation.
    OB gladly acknowledges support by the visitor program of the Max-Planck-Institut für Physik komplexer Systeme.
    
    While completing this work we learned about a similar, independent result by Qi, Gu and Yao who also realized the \textsc{DSem} universality class in various \textsc{QDM}s.\cite{Qi:2014vh}
    We also learned about independent work by Iqbal, Poilblanc and Schuch who introduced semionic \textsc{RVB} states.\cite{Iqbal:2014uf}
\end{acknowledgments}

\bibliographystyle{apsrev4-1}
\bibliography{refs}

\end{document}